\documentclass[letter]{aa}
\usepackage{txfonts}
\usepackage{graphicx}
\usepackage{xcolor}

\def\H2{H$_2$}

\begin{document}

\title{Seeds of Life in Space (SOLIS) III.
Formamide in protostellar shocks: evidence for gas-phase formation}
\author{C. Codella \inst{1} 
 \and 
C. Ceccarelli \inst{2,1} 
\and P. Caselli\inst{3}
\and N. Balucani\inst{4,1}
\and V. Barone\inst{5}
\and F. Fontani\inst{1}
\and B. Lefloch\inst{2}
\and L. Podio\inst{1}
\and S. Viti\inst{6}
\and S. Feng\inst{3}
\and R. Bachiller\inst{7}
\and E. Bianchi\inst{1,8}
\and F. Dulieu\inst{9}
\and I. Jim\'enez-Serra\inst{10}
\and J. Holdship\inst{6}
\and R. Neri\inst{11}
\and J. Pineda\inst{3}
\and A. Pon\inst{12}
\and I. Sims\inst{13}
\and S. Spezzano\inst{3}
\and A.I. Vasyunin\inst{3,14}
\and F. Alves\inst{3}
\and L. Bizzocchi\inst{3}
\and S. Bottinelli\inst{15,16}
\and E. Caux\inst{15,16}
\and A. Chac\'on-Tanarro\inst{3}
\and R. Choudhury\inst{3}
\and A. Coutens\inst{6}
\and C. Favre\inst{1,2}
\and P. Hily-Blant\inst{2}
\and C. Kahane\inst{2}
\and A. Jaber Al-Edhari\inst{2,17}
\and J. Laas\inst{3}
\and A. L\'opez-Sepulcre\inst{11}
\and J. Ospina\inst{2}
\and Y. Oya\inst{18}
\and A. Punanova\inst{3}
\and C. Puzzarini\inst{19}
\and D. Quenard\inst{10}
\and A. Rimola\inst{20}
\and N. Sakai\inst{21}
\and D. Skouteris\inst{5}
\and V. Taquet\inst{22,1}
\and L. Testi\inst{23,1}
\and P. Theul\'e\inst{24}
\and P. Ugliengo\inst{25}
\and C. Vastel\inst{15,16}
\and F. Vazart\inst{5}
\and L. Wiesenfeld\inst{2}
\and S. Yamamoto\inst{18}
        }

\institute{
INAF, Osservatorio Astrofisico di Arcetri, Largo E. Fermi 5,
50125 Firenze, Italy
\and
Univ. Grenoble Alpes, CNRS, Institut de
Plan\'etologie et d'Astrophysique de Grenoble (IPAG), 38000 Grenoble, France
\and
Max-Planck-Institut f\"ur extraterrestrische Physik (MPE), 
Giessenbachstrasse 1, 85748 Garching, Germany
\and
Dipartimento di Chimica, Biologia e Biotecnologie, Via Elce di Sotto 8, 06123 Perugia, Italy
\and
Scuola Normale Superiore, Piazza dei Cavalieri 7, 56126 Pisa, Italy
\and
Department of Physics and Astronomy, University College London, Gower Street, London, WC1E 6BT, UK
\and
IGN, Observatorio Astron\'omico Nacional, Calle Alfonso XII, 28004 Madrid, Spain
\and
Dipartimento di Fisica e Astronomia, Universit\`a degli Studi di Firenze, Italy
\and
LERMA, Universit\'e de Cergy-Pontoise, Observatoire de Paris, 
PSL Research University, CNRS, Sorbonne Universit\'e, UPMC, Univ. Paris 06, 95000 Cergy Pontoise, France
\and
School of Physics and Astronomy Queen Mary, University of London, 327 Mile End Road, London, E1 4NS
\and
Institut de Radioastronomie Millim\'etrique, 300 rue de la Piscine, Domaine
Universitaire de Grenoble, 38406, Saint-Martin d'H\`eres, France
\and
Department of Physics \& Astronomy, The University of Western Ontario, 1151 Richmond Street, London, Ontario, N6A 3K7, Canada
\and
Universit\'e de Rennes 1, Institut de Physique de Rennes, Rennes, France
\and
Ural Federal University, Ekaterinburg, Russia
\and
Universit\'e  de Toulouse, UPS-OMP, IRAP, Toulouse, France
\and
CNRS, IRAP, 9 Av. Colonel Roche, BP 44346, 31028 Toulouse Cedex 4, France
\and
University of AL-Muthanna, College of Science, Physics Department, AL-Muthanna, Iraq
\and
Department of Physics, The University of Tokyo, 7-3-1, 
Hongo, Bunkyo-ku, Tokyo 113-0033, Japan
\and
Dipartimento di Chimica "Giacomo Ciamician", Via F. Selmi, 2, 40126, Bologna, Italy
\and
Departament de Qu\'{\i}mica, Universitat Aut\`onoma de Barcelona, 08193 Bellaterra, Catalonia, Spain
\and
The Institute of Physical and Chemical Research (RIKEN), 2-1, Hirosawa, 
Wako-shi, Saitama 351-0198, Japan
\and
Leiden Observatory, Leiden University, 9513, 2300-RA Leiden, The Netherlands
\and
ESO, Karl Schwarzchild Srt. 2, 85478 Garching bei M\"unchen, Germany
\and
Aix-Marseille Universit\'e, PIIM UMR-CNRS 7345, 13397 Marseille, France
\and
Universit\`a degli Studi di Torino, Dipartimento Chimica Via Pietro Giuria 7, 10125 Torino, Italy
\and
Research Center for the Early Universe, The University of Tokyo, 
7-3-1, Hongo, Bunkyo-ku, Tokyo 113-0033, Japan
}

\offprints{C. Codella, \email{codella@arcetri.astro.it}}
\date{Received date; accepted date}

\authorrunning{Codella et al.}
\titlerunning{Formamide in L1157-B1}

\abstract
{Modern versions of the Miller-Urey experiment claim that 
formamide (NH$_2$CHO) could be the starting point for the 
formation of metabolic and genetic macromolecules. 
Intriguingly, formamide is indeed observed 
in regions forming Solar-type stars as well as in external galaxies.} 
{How NH$_2$CHO is formed has been a puzzle for decades: our goal is to contribute
to the hotly debated question of whether formamide is mostly formed via gas-phase or grain surface chemistry.}
{We used the NOEMA interferometer to image NH$_2$CHO towards the L1157-B1 blue-shifted shock, a well
known interstellar laboratory, to study how the 
components of dust mantles and cores released into the gas phase triggers the
formation of formamide.}
{We report the first spatially
resolved image (size $\sim$ 9$\arcsec$, $\sim$ 2300 AU) of formamide emission in a shocked region around a Sun-like
protostar: the line profiles are blueshifted and have a FWHM $\simeq$ 5 km s$^{-1}$. 
A column density of $N_{\rm NH_2CHO}$ = 8 $\times$ 10$^{12}$ cm$^{-1}$,
and an abundance (with respect to H-nuclei) of 4 $\times$ 10$^{-9}$ are derived. 
We show a spatial segregation of formamide with respect to other organic species. 
Our observations, coupled with a chemical modelling analysis, indicate that 
the formamide observed in L1157-B1 is formed by gas-phase chemical process,
and not on grain surfaces as previously suggested.}
{The SOLIS interferometric observations of formamide provide direct evidence 
that this potentially crucial brick of life is efficiently 
formed in the gas-phase around Sun-like protostars.}


\maketitle

\section{Introduction}
 
One of the main open questions in astrochemistry regards the mechanisms
leading to the formation of the so-called interstellar 
complex organic molecules (iCOMs: molecules with at least 6 atoms),
which can be considered as the building blocks of more complex pre-biotic compounds 
(see e.g. Caselli \& Ceccarelli 2012). 
This topic is even more important in regions
around Sun-like protostars which will produce future Solar-like systems.
In particular, modern versions of the Urey-Miller experiment
suggest that formamide (NH$_2$CHO) might be the starting point of
metabolic and genetic species (Saladino et al. 2012).
Intriguingly, formamide is detected in both Galactic high- and low-mass 
star forming regions
(e.g. Turner 1991; Nummelin et al. 1998; 
Halfen et al. 2011; Kahane et al. 2013; 
Mendoza et al. 2014; L\'opez-Sepuclre et al. 2015)
as well as in external galaxies (M\"uller et al. 2013).
Despite being so easily found, it is still hotly debated how this species and other iCOMs 
are formed (e.g. Herbst \& van Dishoeck 2009). 

The two current theories predict formation by reactions in the gas 
phase (e.g. Vasyunin \& Herbst 2013; Balucani et al. 2015; Vasyunin et al. 2017) 
or on interstellar dust grains (e.g. Garrod \& Herbst 2008; Garrod et al. 2008), the latter through 
surface reactions or induced by energetic processing. 
Focusing on formamide, the gas-phase theory proposes that 
it is synthesised by the reaction of formaldehyde (H$_2$CO) 
and amidogen (NH$_2$), as suggested by Barone et al. (2015) and Vazart et al. (2016). 
Various mechanisms have been advanced for the formation of formamide 
on the grain surfaces including the combination of amidogen and formyl 
radical (HCO; Garrod et al. 2008; Jones et al. 2011), the hydrogenation 
of isocyanic acid (HNCO; Mendoza et al. 2014), 
the latter being most likely an inefficient reaction
(Noble et al. 2015; Song et al. 2016), 
and particle/UV photon irradiation of ice mixtures
(e.g. Ka\v{n}uchov\'a et al. 2016; Fedoseev et al. 2016).

From the observational point of view, it is challenging to 
safely assess which formation mechanism is dominating for formamide.
The chemically rich molecular outflow driven by the
L1157-mm Class 0 protostar ($d$ = 250 pc) is a unique region which  
can be used to tackle this question. 
A precessing, episodic jet of matter at supersonic velocity 
emerges from L1157-mm (Gueth et al. 1996; Podio et al. 2016). 
The jet has excavated two main cavities, with apices  
called B1 and B2 (see Fig. 1). In particular, B1 consists of 
a series of shocks (see Sect. 3) caused by different episodes of ejection 
impacting against the cavity wall (Podio et al. 2016), 
the oldest of which (kinematical age $\simeq$ 1100 yr) is also 
the farthest away from the source. 

Previous observations revealed that in B1 the jet impacts caused 
erosion of the grain cores and ices, producing 
large quantities of gaseous SiO ($\sim$ 10$^{-7}$; Gueth et al. 1998), 
H$_2$O ($\sim$ 10$^{-4}$; Busquet et al. 2014),  
and HCOOCH$_3$ ($\sim$ 10$^{-8}$; Arce et al. 2008) among 
other species (see also Lefloch et al. 2017). 
Hence, L1157-B1 provides us with a perfect place to study the 
reactions occurring when previously frozen species are injected 
into the gas, as their relative abundance evolution depends on 
the relative efficiency of the various reactions. 
Previous studies have shown that any variation on the 1000 AU scale, 
as the one probed by our work, is due to the passage of shocks, 
rather than to differences in the composition of pre-existing, 
pre-shocked dust and gas (Benedettini et al. 2012;
Busquet et al. 2014). 
To conclude, within the context of the study of iCOMs, the advantages
of the L1157-B1 laboratory are twofold:  
(i) the source is not directly heated by the protostar, which is
0.08 pc away, and (ii) solid species in dusty icy mantles have been injected
into the gas phase
due to a jet-induced shock and consequently sputtering
(e.g. Bachiller et al. 2001).  
 
L1157-B1 is one of the targets of the
SOLIS\footnote{http://solis.osug.fr/} 
(Seeds Of Life In Space; Ceccarelli et al. submitted, hereafter Paper I) 
IRAM NOEMA 
(NOrthern Extended Millimeter Array) large program 
to investigate iCOM formation   
during the early stages of the star forming process.
In this Letter we report the first high spatial
resolution NH$_2$CHO image 
and comparison
with the acetaldehyde (CH$_3$CHO), which  
allow us to constrain how gas phase chemistry matters for
the formation of NH$_2$CHO.

\section{Observations} 

The L1157-B1 shock was observed at 3 mm 
with the IRAM NOEMA seven-element array
during several tracks in July, October, and November 2015 using both 
the C and D configurations.
The shortest and longest baselines are 19 m and 237 m,
respectively,   
allowing us to recover
emission at scales up to $\sim$ 17$\arcsec$.
The NH$_2$CHO (4$_{\rm 1,4}$--3$_{\rm 1,3}$) line
($E_{\rm u}$ = 13 K, $S\mu^2$ = 49 D$^2$, $A_{\rm ul}$ = 3.7 $\times$ 10$^{-5}$ s$^{-1}$) 
at 81693.45 MHz\footnote{Spectroscopic parameters 
have been extracted from the Cologne Database for Molecular Spectroscopy (M\"uller et al. 2005).}
was observed using 80 MHz backends with a spectral resolution of  
156 kHz ($\sim$ 0.57 km s$^{-1}$).
We recover about 60\%--70\% of the emission observed by   
Mendoza et al. (2014) using the IRAM-30m 
(see Appendix A, Fig. A.1 for the 30-m and NOEMA spectra).
Calibration was carried out following standard procedures,
using GILDAS-CLIC\footnote{http://www.iram.fr/IRAMFR/GILDAS}.
The bandpass was calibrated on 3C454.3, while the absolute
flux was fixed by observing MWC349 and 0524+034, the latter being 
used also to set the gains in phase and amplitude.
The phase rms was $\le$ 50$\degr$,
the typical precipitable water vapor (PWV) was from 10 mm to 40 mm,
and the system temperatures $\sim$ 80--100 K (D) and $\sim$ 150--250 K (C).
The final uncertainty on the absolute flux scale is $\leq$ 15\%.
The rms noise in the 156-kHz channels was 2 mJy beam$^{-1}$.
Images were produced using natural weighting, and restored with a 
clean beam of 5$\farcs$79 $\times$ 4$\farcs$81 (PA = --94$\degr$).  

\section{Results and discussion}

\begin{figure}
\centerline{\includegraphics[angle=0,width=6.5cm]{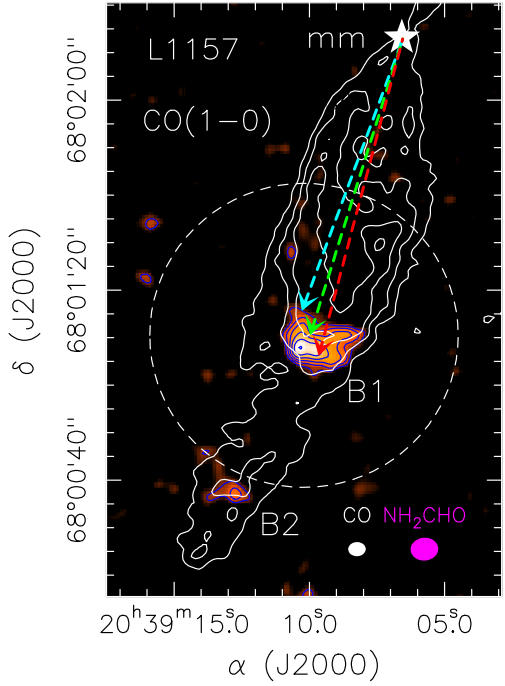}}
\caption{
The L1157 southern blue-shifted lobe
in CO (1--0) (white contours; Gueth et al. 1996). The precessing jet
ejected by the central object L1157-mm (white star) excavated two cavities,
with apices B1 and B2, respectively. The maps are centred at $\alpha({\rm J2000})$ = 20$^h$ 39$^m$ 10$\fs$2,
$\delta({\rm J2000})$ = +68$\degr$ 01$\arcmin$ 10$\farcs$5 ($\Delta$$\alpha$ = +25$\arcsec$ and
$\Delta$$\delta$ = --63$\farcs$5 from L1157-mm).
The emission map of the NH$_2$CHO (4$_{\rm 1,4}$--3$_{\rm 1,3}$) line
(integrated over the velocity range --5 to +5 km s$^{-1}$) is shown by the colour image.
For the CO image, the first contour and step are
6$\sigma$ (1$\sigma$ = 0.5 Jy beam$^{-1}$ km s$^{-1}$) and 4$\sigma$, respectively. The first
contour and step of the NH$_2$CHO map correspond to 3$\sigma$ (15 mJy beam$^{-1}$
km s$^{-1}$) and 1$\sigma$, respectively.
The dashed circle shows the primary
beam of the NH$_2$CHO image (64$\arcsec$). The magenta and white ellipses depict
the synthesised beams of the NH$_2$CHO (5$\farcs$79 $\times$ 4$\farcs$81, PA = --94$\degr$)
and CO (3$\farcs$65 $\times$ 2$\farcs$96, PA=+88$\degr$) observations, respectively.
The three dashed arrows indicate the directions (projected on the plane of the sky) of the
episodic jet producing the shocks analysed in Sect. 3 and Fig. 2.}
\label{maps}
\end{figure}

\subsection{NH$_2$CHO spectra and maps}

Formamide emission has been detected  
towards L1157--B1 with a S/N $\geq$ 8, confirming 
the NH$_2$CHO identification  
done by Mendoza et al. (2014) in the context 
of the ASAI IRAM 30-m spectral survey. 
Figure 1 shows the map of the
NH$_2$CHO (4$_{\rm 1,4}$--3$_{\rm 1,3}$) integrated emission on
top of the CO (1--0) image (Gueth et al. 1996), which well outlines
the B1 and B2 cavities opened by
the precessing jet driven by the L1157-mm protostar, located
(in Fig. 1) at $\Delta$$\alpha$ = --25$\arcsec$ and
$\Delta$$\delta$ = +63$\farcs$5. 
Formamide is emitted from an extended region
with a beam deconvolved size of $\simeq$ 9$\arcsec$ ($\sim$ 2300 AU) 
which is clearly associated with the  
apex of the B1 cavity. In addition, weaker (S/N $\geq$ 4) emission appears
in correspondence with the older B2 peak, which, however, being more than
30$\arcsec$ from B1 is affected by primary-beam attenuation. 
For that reason, the further analysis will be focused on the B1 region.
The line at the peak emission (see Fig. A.1) has a linewidth with a 
Full Width Half Maximum (FWHM) of 4.6$\pm$0.6 km s$^{-1}$ and peaks
close to $\sim$ 0 km s$^{-1}$, thus being
blueshifted ($v_{\rm sys}$ = +2.6 km s$^{-1}$; e.g. Bachiller et al. 2001).
Using the emitting size and assuming optically thin conditions and an excitation temperature of 10 K 
(as derived by several formamide lines observed with the IRAM 30-m antenna;
Mendoza et al. 2014), the average formamide column density is 
$N_{\rm NH_2CHO}$ = 8 $\times$ 10$^{12}$ cm$^{-1}$. 
This corresponds to an estimated average abundance (with respect to H-nuclei) of about 
4 $\times$ 10$^{-9}$, assuming a H column density of 2 $\times$ 10$^{21}$ cm$^{-2}$ 
derived for the cavity by Lefloch et al. (2012).

\subsection{Formamide and acetaldehyde spatial anticorrelation}

\begin{figure}
\centerline{\includegraphics[angle=0,width=8.5cm]{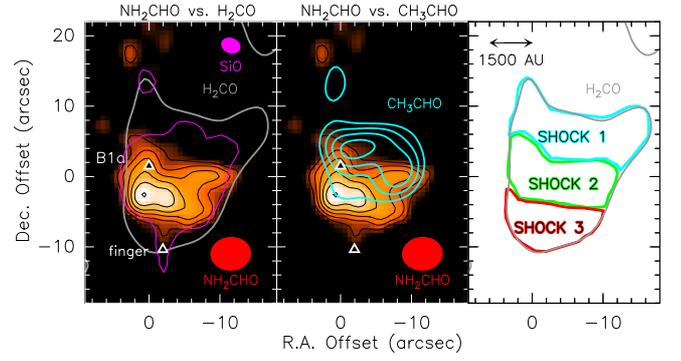}}
\caption{Chemical segregation in L1157-B1.  
The maps are centred at:
$\alpha({\rm J2000})$ = 20$^h$ 39$^m$ 09$\fs$5,
$\delta({\rm J2000})$ = +68$\degr$ 01$\arcmin$ 10$\farcs$0. 
{\it Left panel:}  p-H$_2$CO (2$_{0,2}$--1$_{0,1}$) integrated emission 
(grey; Benedettini et al. 2013), and SiO (2--1) (magenta) at low-velocity 
(less than --10 km s$^{-1}$ with respect to $v_{\rm sys}$; Gueth et al. 1998) 
on top of the present NH$_2$CHO line emission map (colour image, black contours). 
For clarity, for the H$_2$CO and SiO images, only the 3$\sigma$ 
(1$\sigma$ = 3.3 mJy beam$^{-1}$ km s$^{-1}$ and 50 mJy beam$^{-1}$ km s$^{-1}$ for H$_2$CO and SiO, 
respectively) contour is reported to show the overall B1 structure (see Fig. A.2 
for the complete set of contours). 
The northern triangle at “B1a” identifies the youngest position in B1, where 
the precessing jet driven by L1157-mm impacts the cavity wall 
(Gueth et al. 1998; Busquet et al. 2014), while the southern triangle denotes the position 
of the oldest shock, which is identified by the so-called "finger" feature 
traced by SiO at low-velocity. The H$_2$CO image is smoothed to the same angular 
resolution (red ellipse; see Fig. 1) of the NH$_2$CHO (4$_{\rm 1,4}$--3$_{\rm 1,3}$) line data. 
The synthesised beam of the SiO is: 2$\farcs$8 $\times$ 2$\farcs$2, PA = 56$\degr$.  
{\it Middle panel:} Same as in the left panel for the CH$_3$CHO (7$_{0,7}$--6$_{0,6}$ E+A) 
velocity-integrated emission (cyan contours; also smoothed to the same beam of the NH$_2$CHO map; first contour and step 
correspond to 3$\sigma$, 4 mJy beam$^{-1}$ km s$^{-1}$, and 1$\sigma$, respectively;
Codella et al. 2015). 
{\it Right panel:} Sketch of the three zones identified from the spatial distribution 
of formamide and acetaldehyde: SHOCK 1 (blue), the northern region, where CH$_3$CHO 
(and not NH$_2$CHO) is detected; SHOCK 2 (green), where both CH$_3$CHO and NH$_2$CHO 
are detected; and SHOCK 3 (red), the southern region, where only NH$_2$CHO (and not CH$_3$CHO) 
is detected. Time increases and chemistry evolves going from SHOCK 1 to SHOCK 3 
(see dashed arrows in Fig. 1).} 
\label{zoom}
\end{figure}

In Figure 2, we report a zoom-in of the B1 structure, as traced by the line 
emission from formaldehyde and SiO (Gueth et al. 1998; Benedettini et al. 2013). 
The figure clearly shows the first important 
result of these observations: the formamide emission does not coincide with that from 
H$_2$CO and SiO, but only covers the southern portion of the B1 structure. 
Also, Fig. 2 reports the emission from another iCOM, acetaldehyde (CH$_3$CHO; Codella et al. 2015). 
Surprisingly, unlike formamide,  
it is mostly associated with the northern portion of B1. When the difference between 
these two species is considered, one can identify three zones as follows:
SHOCK 1: the northern and youngest one, where only acetaldehyde emits
($X$(CH$_3$CHO)/$X$(NH$_2$CHO) abundance ratio $>$ 8; see Fig. 3); 
SHOCK 2: an intermediate zone, where both formamide and acetaldehyde are present
($X$(CH$_3$CHO)/$X$(NH$_2$CHO) = 2--8); 
SHOCK 3: the southern and oldest region, where only formamide emits
($X$(CH$_3$CHO)/$X$(NH$_2$CHO) $<$ 2). 

The analysis of the SiO and HDCO distribution (Figure A.2) confirms that
B1 is composed by at least two different shocks  
and is not a single bow-like shock. Specifically:
(A) The northern part, SHOCK 1, is associated with the youngest shock (within the B1 structure) 
at B1a, which is characterised by (i) the emission of HDCO, a selective 
tracer of dust mantle release (Fontani et al. 2014), and (ii) extremely 
high-velocity SiO emission, tracing the current sputtering of the dust refractory 
cores. 
(B) The southern region, SHOCK 3 is associated with the oldest shock (within B1), 
because (i) no HDCO is observed, and (ii) SiO emission is only observed 
at low-velocity and shows a “finger” pointing South (Gueth et al. 1998). 
This implies that either SiO molecules have been slowed down with time with respect 
to the high-velocities (needed to produce gaseous Si) or that the shock incident 
angle has changed, so that the projected velocity is lower. In both cases this 
indicates that a unique shock structure for L1157-B1 is ruled out.
(C) The central region, between SHOCK 1 and SHOCK 3, is characterised by 
the occurrence of the bulk of the low-velocity SiO molecules, which, once produced 
at high-velocities, have plausibly slowed down with time. 
It is then reasonable to assume that this region is associated with a third, 
intermediate in time, shock event. Note however, that the results and 
conclusions of the present paper are based on the occurrence of at least two 
shocks of different age (SHOCKs 1 and 3). 

We notice that the difference in the three zones cannot be attributed to excitation effects, 
as the mapped formamide and acetaldehyde lines have similar upper level energies (11 K and 26 K), 
similar Einstein coefficients ($\sim$ 10$^{-5}$ s$^{-1}$), and the derived excitation 
temperatures are also similar (10 K against 15 K, for formamide and acetaldehyde 
respectively, Mendoza et al. 2014; Codella et al. 2015). Besides, there is 
no evidence of a monotonic volume density gradient across the B1 region from North to South
(Benedettini et al. 2013; G\'omez-Ruiz et al. 2015), as in the case of CH$_3$CHO/NH$_2$CHO line intensity ratio 
(see Fig. A.2). Therefore, 
the difference between the three zones must be due to a difference in the chemical 
composition, thus indicating a clear evolutionary effect.

\section{Chemical modelling}

To understand what the observed chemical differentiation implies, we 
ran an astrochemical model (a modified version of Nahoon, Loison et al. (2014), see Appendix B)
considering 
three possibilities: (i) formamide and acetaldehyde are grain-surface chemistry products, 
(ii) formamide and acetaldehyde are gas-phase chemistry products, (iii) one of the two species 
is a grain-surface and the other one a gas-phase chemistry product. Briefly, we use a time-dependent 
gas-phase code that follows the chemical evolution of the gas. It starts with the chemical composition 
of a molecular cloud and then simulates the passage of the shock by suddenly increasing the 
gas density and temperature (to 10$^5$ cm$^{-3}$ and 60 K, respectively,
i.e. typical values measured for the B1 cavities: 20--80 K, Lefloch et al. 2012; 0.5--10 $\times$ 10$^5$ cm$^{-3}$; G\'omez-Ruiz et al. 2015), and the gaseous abundance of 
grain mantle molecules. The abundances of the mantle molecules are assumed to be similar to those measured 
by IR observations of the dust ices (Boogert et al. 2005) or specifically constrained by previous 
studies on L1157-B1 (see Table B.1). The chemical network is described in Appendix B.
The results of the modelling are discussed for the three cases mentioned above. 

\begin{figure}
\centerline{\includegraphics[angle=0,width=10cm]{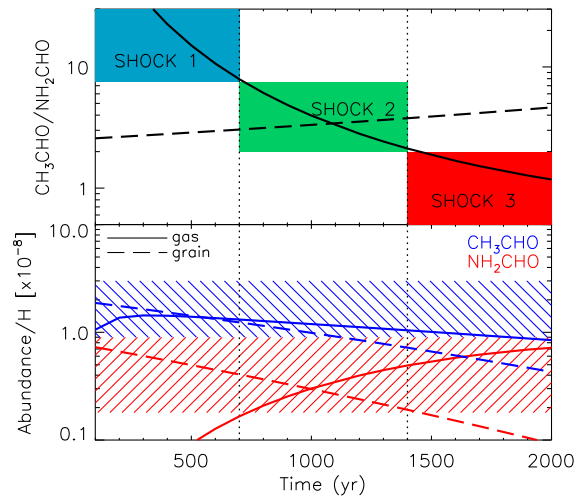}}
\caption{
Model predictions of acetaldehyde and formamide after a shock passage.
{\it Upper panel:} CH$_3$CHO/NH$_2$CHO calculated abundance ratio as a function
of time from the passage of the shock. Solid lines refer to a model where acetaldehyde and
formamide are both synthesized in the gas phase, whereas dashed lines refer to predictions
assuming that they are injected into the gas phase directly from the grain mantles.
The vertical ranges of the three coloured boxes represent the measured ranges,
including the uncertainties, of the CH$_3$CHO/NH$_2$CHO abundance ratio towards
the three zones identified in Fig. 2: SHOCK 1 (ratio $>$ 8), SHOCK 2 (ratio 2--8),
and SHOCK 3 (ratio $\leq$ 2). When not detected, we derived the upper limits
on CH$_3$CHO and NH$_2$CHO by using the 3$\sigma$ value. The two dotted
vertical lines define the time when the formamide and acetaldehyde abundance ratios
(as derived by the gas-phase model) fall below the minimum measured values.
{\it Lower panel:} Acetaldehyde (CH$_3$CHO, blue) and formamide (NH$_2$CHO, red) abundances,
with respect to H nuclei, as a function of time from the passage of the shock.
The dashed blue and red regions show the maximum and minimum CH$_3$CHO and NH$_2$CHO
measured abundances.}
\label{fig-model}
\end{figure}

\noindent
{\it 1. Grain-surface formation of CH$_3$CHO and NH$_2$CHO}:
First, we assume that both formamide and acetaldehyde are synthesised on the grain surfaces and 
that the passage of the shock injects these two species into the gas phase in quantities 
such that the measured abundances are roughly reproduced. The predicted abundances as 
a function of time are shown in Figure 3. They decrease by approximately the same factor in a 2000 yr interval. 
Actually, the predicted [CH$_3$CHO]/[NH$_2$CHO] abundance ratio slightly increases with time, 
which is in contrast with the observations that show exactly the opposite trend. Therefore, 
the pure grain-surface hypothesis cannot explain the observed formamide/acetaldehyde segregation. 
In other words, our observations rule out the hypothesis that the bulk of the observed acetaldehyde 
and formamide in L1157-B1 are both directly injected from the grain mantles into the gas phase.

\noindent
{\it 2. Gas-phase formation of CH$_3$CHO and NH$_2$CHO}:
We then assumed that both acetaldehyde and formamide are formed in the gas phase from species 
previously on the grain mantles and injected into the gas-phase during the shock passage. 
Acetaldehyde is assumed to be formed by the reaction of ethyl radical (CH$_3$CH$_2$) with atomic oxygen
(Charnley et al. 2004): CH$_3$CH$_2$ + O $\to$ CH$_3$CHO + H. 
Formamide is assumed to be formed by the reaction of amidogen with formaldehyde 
(Barone et al. 2015; Vazart et al. 2016): 
NH$_2$ + H$_2$CO $\to$ NH$_2$CHO + H.  We run various models with different values of ethyl radical, 
ammonia (mother of NH$_2$), and formaldehyde, to reproduce the observed abundances. We also run 
alternative tests injecting ethane (CH$_3$CH$_3$), the fully hydrogenated ``cousin'' of ethyl radical, 
and amidogen, a partially hydrogenated ``cousin'' of ammonia, into the gas. The best agreement 
with observations is obtained by injecting into the gas phase 4 $\times$ 10$^{-8}$ of ethyl radical, 
2 $\times$ 10$^{-5}$ of ammonia and 1 $\times$ 10$^{-6}$ of formaldehyde (see Appendix B for details). 
This model not only reproduces fairly well the observed abundances (see Fig. 3), it also fits the behaviour of the 
[CH$_3$CHO]/[NH$_2$CHO] abundance ratio, with acetaldehyde being more abundant in the younger northern 
SHOCK 1 and formamide being more abundant in the older southern SHOCK 3. 
Note that the evolution timescale is sensitive to the cosmic ray ionisation rate $\zeta$. 
We find that the best agreement with the observations is obtained 
when $\zeta$ is 6 $\times$ 10$^{-16}$ s$^{-1}$, which is very close to that previously found
(Podio et al. 2014), based on the analysis of the molecular ions in L1157-B1.  
Finally, a larger shocked gas density would result in speeding up 
the chemical evolution. As a consequence, the CH$_3$CHO/NH$_2$CHO 
abundance ratio curve would be shifted towards earlier times. 
For example, if the density were ten times larger, namely 
$2 \times 10^6$ cm$^{-3}$, the curve would be shifted earlier 
by about 1000 yr. This just means that a substantial difference, 
by a factor ten, in the gas density at SHOCKs 1 and 3 would not 
change our major conclusions, but would just imply a slightly 
smaller cosmic ray ionisation rate.

\noindent
{\it 3. Either acetaldehyde or formamide is a grain-surface and the other a gas-phase chemistry product:} 
We checked the possibility that acetaldehyde is synthesised on the grain surfaces and formamide in the 
gas and we obtained results similar to the case (2). Hence, the gaseous CH$_3$CHO abundance evolution 
is rather independent on the formation route (surface chemistry or gas-phase chemistry). 
We finally checked the possibility that the gas-phase reaction NH$_2$ + H$_2$CO is not efficient
(Song \& K\"astner 2016). In this case, no model can reproduce the observations (both the abundance and the evolution). 

In summary, the new SOLIS observations indicate that the formation of observed formamide in L1157-B1 
is dominated by gas-phase reactions involving species previously hydrogenated on the grain surfaces, although 
we cannot exclude a minor contribution from mechanisms such as energetic processing of ices. 
The formamide abundance needs to peak when the acetaldehyde abundance has already started to decrease. 
This is only possible if formamide is mostly formed in the gas phase and the reaction between 
amidogen and formaldehyde (Barone et al. 2015; Vazart et al. 2016) successfully reproduces the observations. 
Although simple, our model catches the essential aspects of the chemical behaviour of formamide and 
acetaldehyde, namely their abundance as a function of time once the shock has passed. 
Indeed, the major uncertainties lie in the used chemical network more than the detailed physical processes or 
the detailed gas-grain interactions (see Appendix B for more). In this context, it is encouraging that 
the age difference between SHOCK 1 and SHOCK 3 derived by our simple astrochemical model ($\sim$ 700 yr) is 
the same order of magnitude of the one ($\sim$ 2000 yr) independently derived by dynamical studies 
of L1157-B1 (Podio et al. 2016). A more detailed modelling including a more complex 
and realistic treatment of the shock will be necessary to confirm that this is not just a coincidence 
and to refine the present predictions. 

\section{Conclusions}

The present work demonstrates that the formamide observed in L1157-B1 is 
dominated by gas-phase chemistry and that the reaction NH$_2$ + H$_2$CO $\to$ NH$_2$CHO + H 
explains the observations. Although we are unable to place constraints
on the acetaldehyde formation route, we note that quantum chemistry computations have
shown that the simple combination of the methyl radical (CH$_3$) and formyl radical (HCO) is an
inefficient channel on water ice surfaces (Enrique-Romero et al. 2016), so
that it is possible that CH$_3$CHO is also a gas-phase product. 
The recent detection of iCOMs in cold objects 
(e.g. Vastel et al. 2014) has already challenged a pure grain-surface 
chemistry paradigm for their formation (e.g. Vasyunin \& Herbst 2013, and
references therein). These new observations 
add evidence that gas-phase chemistry plays an important role in the game 
of iCOM formation.

\begin{acknowledgements}
We are very grateful to all the IRAM staff, whose dedication 
allowed us to carry out the SOLIS project. 
We also thank the anonymous referee for useful suggestions.
This work was supported 
by (i) the French program “Physique et Chimie du Milieu Interstellaire” 
(PCMI) funded by the Conseil National de la Recherche Scientifique 
(CNRS) and Centre National d'Etudes Spatiales (CNES), (ii) by the 
Italian Ministero dell'Istruzione, Universit\`a e Ricerca through 
the grant Progetti Premiali 2012 - iALMA (CUP C52I13000140001), 
(iii) by the program PRIN-MIUR 2015 STARS in the CAOS - Simulation Tools 
for Astrochemical Reactivity and Spectroscopy in the Cyberinfrastructure 
for Astrochemical Organic Species (2015F59J3R, MIUR Ministero 
dell'Istruzione, dell'Universit\`a della Ricerca e della 
Scuola Normale Superiore), and (iv) by the French Agence Nationale de 
la Recherche (ANR), under reference ANR-12-JS05-0005. PC, ACT, JEP, 
and APu acknowledge support from the European Research Council 
(ERC; project PALs 320620).  APo acknowledges the financial support 
provided by a Canadian Institute for Theoretical Astrophysics (CITA) 
National Fellowship. I.J.-S. and D.Q. acknowledges the financial support 
received from the STFC though an Ernest Rutherford Fellowship 
(proposal number ST/L004801) and Grant (ST/M004139).
\end{acknowledgements}

\clearpage

\appendix

\section{Additional line spectra and maps}

Figure A.1 (Upper panel) shows the comparison in flux density scale 
between the NH$_2$CHO (4$_{\rm 1,4}$--3$_{\rm 1,3}$) spectrum 
as observed using the IRAM 30-m antenna (Mendoza et al. 2014) 
and that extracted from the present NOEMA map from 
a circular region equal to the IRAM 30-m Half Power Beam Width (HPBW) 
of 30$\arcsec$. The lines are blue-shifted ($v_{\rm sys}$ = +2.6 km s$^{-1}$;
Bachiller et al. 2001). Between 60\% and 70\% of the emission 
observed using the IRAM single-dish is recovered by the NOEMA 
interferometer, which filters out emission structures 
larger than 17$\arcsec$. Figure A.1 (Bottom panel) also shows the 
NH$_2$CHO (4$_{\rm 1,4}$--3$_{\rm 1,3}$) emission line 
(in brightness temperature scale, T$_{\rm B}$) 
observed at the peak of the formamide spatial distribution (Figs. 1, 2).

\begin{figure}
\centerline{\includegraphics[angle=0,width=4.5cm]{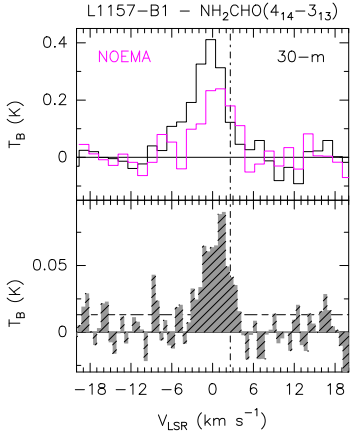}}
\caption{{\it Upper panel:} Comparison in $T_{\rm B}$ scale 
of the NH$_2$CHO (4$_{\rm 1,4}$--3$_{\rm 1,3}$)
spectrum as observed using the IRAM 30-m antenna (Mendoza et al. 2014) and that
extracted from the present NOEMA map from a circular region equal to the IRAM 30-m HPBW (30$\arcsec$). $F_{\rm \nu}$(Jy) = 4.9493 $T_{\rm B}$(K). The NOEMA spectrum has been smoothed 
to match the IRAM 30-m velocity resolution.
{Bottom panel:} emission (in $T_{\rm B}$ scale) extracted at
the peak of the formamide spatial distribution (see Fig. 2). Horizontal dashed line 
indicates the 1$\sigma$ noise level (13 mK).}
\label{spectra}
\end{figure}

Figure A.2 shows how different shocks are present within the L1157-B1 structure.
The northern part (see the B1a position) is associated with both SiO emitting  
very high velocities (up to --18 km s$^{-1}$
with respect to $v_{\rm sys}$), as well as with HDCO, a selective tracer of dust mantle release.
On the other hand, the sourhern region is characterised by no HDCO, and by low-velocity 
SiO emission producing the so-called ``finger'' pointing towards South (Gueth et al. 1998).

\begin{figure}
\centerline{\includegraphics[angle=0,width=8.5cm]{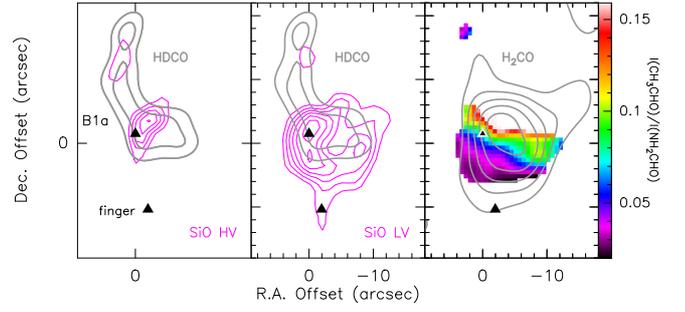}}
\caption{Different shocks in L1157-B1, as traced by HDCO and SiO high- and low-velocity emission. 
The maps are centred as in Fig. 2.
The beams of all the images are reported in Figs. 1 and 2. 
{\it Left panel}: HDCO (2$_{1,1}$--1$_{0,1}$) integrated emission (grey contours; 
smoothed to the same beam of the NH$_2$CHO map; Fontani et al. 2014) 
tracing the dust mantle release due to the youngest (within B1) impact (the B1a position; 
northern black triangle) of the jet against the cavity wall (Gueth et al. 1998; Busquet et al. 2014). 
The magenta contours represent the high-velocity (HV; from --10 km s$^{-1}$ up to --18 km s$^{-1}$ 
with respect to $v_{\rm sys}$) SiO (2--1) emission, tracing a smaller region 
associated with the release of SiO from the dust refractory core. For the HDCO image, 
the first contour and step are 3$\sigma$ (1$\sigma$ = 1 mJy beam$^{-1}$ km s$^{-1}$) and 
1$\sigma$, respectively. For SiO at HV, the first contour and step are 5$\sigma$ 
(1$\sigma$ = 27 mJy beam$^{-1}$ km s$^{-1}$) and 10$\sigma$, respectively. 
{\it Middle panel:} Same as in the left panel for HDCO. The magenta contours represent the 
low-velocity (LV; up to --10 km s$^{-1}$ with respect to $v_{\rm sys}$) SiO (2--1) emission, 
tracing a large structure extending towards the south (up to 16$\arcsec$, 4000 AU, from B1a), 
which creates the so-called ``finger'' feature (southern black triangle; Gueth et al. 1998) 
that coincides with the location of the oldest shock within B1. For SiO at LV, the first 
contour and step are 5$\sigma$ (1$\sigma$ = 50 mJy beam$^{-1}$ km s$^{-1}$) and 10$\sigma$, 
respectively. {\it Right panel:} p-H$_2$CO (2$_{0,2}$--1$_{0,1}$) integrated emission 
(grey; also smoothed to the same beam of the NH$_2$CHO image; Benedettini et al. 2013). 
The first contour and step are 3$\sigma$ (1$\sigma$ = 
3.3 mJy beam$^{-1}$ km s$^{-1}$) and 6$\sigma$, respectively. In colour 
is reported the CH$_3$CHO (7$_{0,7}$--6$_{0,6}$ E+A)/NH$_2$CHO (4$_{1,4}$--3$_{1,3}$) line 
intensity ratio (derived where both emission is at least 3$\sigma$) smoothly decreasing 
from North to South (see the wedge).}
\label{maps2}
\end{figure}

\section{Sensitivity to the model parameters}

In order to understand the origin of the observed spatial segregation between the acetaldehyde 
and formamide emission, we ran a chemical model with the aim to simulate the passage of the shock. 
To this end, we used a modified (to make it more flexible) version of Nahoon (Loison et al. 2014), and a chemical 
network consisting of 511 species and 7792 reactions. The base of the chemical network is KIDA.2014 
(http://kida.obs.u-bordeaux1.fr), which has been augmented and corrected with updated reactions 
(Loison et al. 2014; Balucani et al. 2015; Barone et al. 2015). To simulate the passage of the shock, 
we followed the strategy used in previous works (Podio et al. 2014; Codella et al. 2015), namely a 2-step modelling. 
In the first step, we ran a model assuming the conditions of the gas before the passage of the shock, 
namely a gas cloud of 2 $\times$ 10$^4$ H-nuclei cm$^{-3}$ and a temperature of 10 K. The cosmic-ray 
ionisation rate was previously constrained to be $\sim$ 3 $\times$ 10$^{-16}$ s$^{-1}$ (Podio et al. 2014). 
The steady state abundances are then used as initial abundances for modelling the second step, 
with exceptions of the species that are injected into the gas phase because of the shock passage. 
In this second step, the density is set at 2 $\times$ 10$^5$ H-nuclei cm$^{-3}$ and the temperature at 60 K
(Lefloch et al. 2012; G\'omez-Ruiz et al. 2015).  The shock passage is accompanied by the 
sputtering of several species from the grain mantles into the gas phase, which corresponds to 
a sudden increase of their abundance. Again, following previous works, we increased the 
abundances of these gaseous species to simulate the sputtering. Table B.1 lists the species 
injected into the gas and their assumed abundances. The injected species have abundances similar to those measured 
by IR observations of the interstellar dust ices (Boogert et al. 2015). Specifically, they were 
constrained in order to match the abundances derived through direct observations of the 1100 years 
old L1157-B1a shock (Tafalla et al. 1995; Benedettini et al. 2013; Busquet et al. 2014). 
For some injected species, we choose the values derived by comparison of observations with model predictions. 
In addition, we slightly changed the values to fit the observed acetaldehyde and formamide abundances. 
Please note that the final model also reproduces the observed abundances of the species reported in Table B.1. 
When possible, we report the observed gas-phase abundances towards L1157-B1a, i.e. the youngest (1100 yr) 
shocked region within the B1 structure.
Finally, we note that we use a pure gas-phase model with no freeze-out included, since the involved timescale 
is too short for freeze-out to have any impact on the results.
In the following, we give details on the first two cases (the third one is a combination of both) discussed in the main text:

\begin{table}
\caption{Abundances (with respect to H-nuclei) of the species injected 
into the gas in the second step of the model, and previously observed towards L1157-B1a.}           
\begin{tabular}{lcccc}
\hline
\multicolumn{1}{c}{Species} &
\multicolumn{1}{c}{Injected (/H)} &
\multicolumn{2}{c}{Observed (/H)} &
\multicolumn{1}{c}{Reference} \\
\multicolumn{1}{c}{ } &
\multicolumn{1}{c}{} &
\multicolumn{1}{c}{Ices} &
\multicolumn{1}{c}{L1157-B1} &
\multicolumn{1}{c}{ } \\   
\hline
CO$_{2}$ & 3 $\times$ 10$^{-5}$ & $\leq$ 3 $\times$ 10$^{-4}$ & -- & 1 \\
H$_{2}$O & 2 $\times$ 10$^{-4}$ & 1.2 $\times$ 10$^{-4}$ & 1--3 $\times$ 10$^{-4}$  & 1,2 \\
OCS & 2 $\times$ 10$^{-6}$ & -- & $\simeq$ 10$^{-6}$$^a$  & 3 \\
CH$_3$OH & 4 $\times$ 10$^{-6}$ & -- & 5.5 $\times$ 10$^{-6}$$^a$  & 4 \\
H$_2$CO & 1 $\times$ 10$^{-6}$ & -- & 1.5 $\times$ 10$^{-6}$  & 4 \\
NH$_3$ & 2 $\times$ 10$^{-5}$ & -- & $>$ 10$^{-6}$  & 5 \\
CH$_3$CH$_2$ & 4 $\times$ 10$^{-8}$ & -- & 2 $\times$ 10$^{-7}$$^a$  & 6 \\
\hline
\end{tabular}

1. Boogert et al. (2015); 2. Busquet et al. (2014); 3. Podio et al. (2014); 4. Benedettini et al. (2013);
5. Tafalla \& Bachiller (1995); 6. Codella et al. (2015). 
$^a$ Note that these values have been indirectly derived by comparison of observations and model predictions. \\ 
\end{table}

(1) Acetaldehyde and formamide are grain-surface chemistry products: the hypothesis is that 
both species are injected into the gas phase directly from the grain mantles, {\it regardless of the mechanism 
that form them there}. Once in the gas, the two species undergo reactions that destroy them (Fig. 3).  
Specifically, both acetaldehyde and formamide are attacked by the most abundant gaseous ions, namely 
H$_3$O$^+$, H$_3$$^+$ and HCO$^+$, which produce protonated acetaldehyde and protonated formamide, respectively. 
Protonated formamide rapidly recombines with electrons and forms back formamide in only 20\% of cases, 
according to the KIDA database (formamide is not present in the UMIST database). Similarly, 
the recombination of protonated acetaldehyde produces acetaldehyde in only 9\% of electron recombinations. 
While the rate and 
products of the protonated formamide recombination are guessed, those of acetaldehyde, from the UMIST database, 
are measured (Hamberg et al. 2010). 
However, since the two species are destroyed by the same ions, even if the branching ratios 
of the formamide recombination in the KIDA database are wrong, what matters is the percentage of 
electron recombinations that give back formamide, which is certainly not unity. In this respect, therefore, 
the result that formamide and acetaldehyde are not both grain-surface chemistry products is robust.
(2) Acetaldehyde and formamide are gas-phase chemistry products: the hypothesis is that both species 
are produced by gas-phase reactions after the injection 
into the gas of species previously frozen on the grain mantles. 
Only one reaction is known for the gas-phase formation of formamide, NH$_2$ + H$_2$CO $\to$ NH$_2$CHO + H 
(Barone et al. 2015; Vazart et al. 2016; Skouteris et al. 2017). The two mother species injected 
from the grain mantles to synthesise formamide are formaldehyde and ammonia. Both species have been detected 
in the solid state (Boogert et al. 2015) and are thought to be the result of hydrogenation on 
the grain surfaces of CO and N, 
respectively. NH$_2$ is then produced from ammonia via the reactions of NH$_3$ with H$_3$O$^+$ and H$_3$$^+$, 
which both give protonated ammonia NH$_4$$^+$. The electron recombination of NH$_4$$^+$ then produces amidogen. 

For acetaldehyde, a dozen reactions are listed in the KIDA and UMIST databases. Among them, 
the reaction O + CH$_3$CH$_2$ $\to$ CH$_3$CHO + H (Charnley et al. 1992; Harding et al. 2005; Yang et al. 2005)
is the most efficient in the conditions appropriate for the L1157-B1 gas. Therefore, the two gaseous 
species necessary to synthesise acetaldehyde are atomic oxygen and ethyl radical. According to the 
astrochemical models, less than 20\% of gaseous oxygen is in the form of atomic oxygen. 
In L1157-B1, the O abundance is predicted to be 5 $\times$ 10$^{-6}$, in agreement with 
the bright [OI]-63 $\mu$m line observed (Benedettini et al. 2012) by Herschel in L1157-B1. The case of ethyl radical 
is a bit more complicated. It may be the result of the partial hydrogenation of C$_2$H$_2$ or C$_2$H$_4$ 
on the grain surfaces (to be noted that no observations exist about the abundance of this 
species in the solid form and no computations have been carried out) and be directly sputtered 
from the grain mantles as such. On the other hand, the ethyl radical can be produced starting 
from ethane, which, in turn, is formed by the total hydrogenation of C$_2$H$_2$ or C$_2$H$_4$ on 
the ice before sputtering. We run, therefore, a case where only ethane is liberated into the gas-phase: 
even assuming an injection of 4 $\times$ 10$^{-6}$ ethane, namely $\sim$ 10\% of CO, the predicted 
acetaldehyde abundance remains ten times lower than the observed one. Therefore, ethyl radical 
needs to be directly injected into the gas phase from the grain mantles. As discussed in the 
main text, an abundance of 4 $\times$ 10$^{-8}$ is necessary to reproduce the L1157-B1 observations. 
Assuming an ethane abundance of 4 $\times$ 10$^{-6}$, this would imply that about 1\% of 
it is liberated from the grain mantles as the partially hydrogenated ``cousin'' ethyl radical. 
Alternatively, it is possible that the full hydrogenation leading to ethane is not very 
efficient on the grain surfaces.

In order to test the robustness of the results showed in Fig. 3 and discussed in the main text, 
we also run a case where 1\% of ammonia is directly injected as amidogen (namely 2 $\times$ 10$^{-7}$), 
in (possible) analogy to the ethyl radical. In this case, we obtain almost exactly the same 
results shown in Fig. 3, with differences of a few \% within the first 2000 yr, confirming 
that the important mother species in the formamide formation is indeed ammonia.
To summarise, the comparison between the observations and the model predictions leads to ammonia and 
ethyl radical as being the two {\it needed} previously frozen mother species of 
acetaldehyde and formamide respectively, with frozen-and-injected amidogen and ethane being minor actors.

Finally, it is possible that before equilibrating at 60 K, the shocked gas passed through a high-temperature period. 
In order to verify whether this period would affect the results reported in Fig. 3 and our conclusions, 
we run two models with the gas temperature equal to 1000 K, in the case (1) and the case (2). We found:
(1) Acetaldehyde and formamide are grain-surface chemistry products: 
during the first 2000 yr of a possible high-temperature period the predicted abundance ratio of acetaldehyde 
and formamide remains practically the same, as they are destroyed by the same molecular ions (H$_3$O$^+$, 
H$_3$$^+$, and HCO$^+$), so that it does not affect the output of Fig. 3 and 
our conclusion that this case does not reproduce the observed behaviour.
(2) Acetaldehyde and formamide are gas-phase chemistry products: at 1000 K, the formamide rate 
of formation in the gas is very low, as it decreases with a power of 2.56 in temperature (Vazart et al. 2016), 
so that no formamide is appreciably synthesised during the high-temperature period. 
On the contrary, the predicted acetaldehyde abundance is almost the same as the one at 60 K during 
the first 2000 yr. Therefore, a high-temperature period preceding the 60 K one would go towards the 
same direction of our conclusions: the region with only acetaldehyde is younger (and possibly also warmer), 
while the region with formamide identifies an older shocked region and formamide is synthesised via the 
gas-phase reaction NH$_2$ + H$_2$CO.
In summary, even if a high-temperature period, not included in our simple model, preceded the 
present 60 K equilibrated gas temperature, the effects would not change our main conclusion, 
namely that formamide has to be a gas chemistry product.

\end{document}